\documentclass[prb,twocolumn,superscriptaddress,showpacs,amsmath,amssymb]{revtex4}
\usepackage{amsfonts}
\usepackage{bbm}
\usepackage{graphicx}
\usepackage{dcolumn}
\usepackage{bm}

\begin{document}
\title{Josephson diode based on conventional superconductors and a chiral quantum dot}

\author{Qiang Cheng}
\email[]{chengqiang07@mails.ucas.ac.cn}
\affiliation{School of Science, Qingdao University of Technology, Qingdao, Shandong 266520, China}
\affiliation{International Center for Quantum Materials, School of Physics, Peking University, Beijing 100871, China}

\author{Qing-Feng Sun}
\email[]{sunqf@pku.edu.cn}
\affiliation{International Center for Quantum Materials, School of Physics, Peking University, Beijing 100871, China}
\affiliation{Collaborative Innovation Center of Quantum Matter, Beijing 100871, China}
\affiliation{Beijing Academy of Quantum Information Sciences, West Bld.\#3,No.10 Xibeiwang East Rd., Haidian District, Beijing 100193, China}

\begin{abstract}
We propose theoretically a Josephson diode consisting of the conventional superconductors with the plain $s$-wave pairing and a chiral quantum dot.
When an external magnetic field is exerted on the quantum dot, the critical current of the Josephson structure is different for the opposite directions of current flow. The strong nonreciprocity can be obtained in a large area of the parameter space.
The inversion and the on/off state of the nonreciprocity can be conveniently regulated by adjusting the direction of the external field.
The dependences of the nonreciprocal behaviors on the the hopping amplitude, the magnitude of the magnetic field and the energy level of the quantum dot are investigated in details using the Keldysh nonequilibrium Green's function formalism under the self-consistent procedure.
The symmetric and the antisymmetric properties of the nonreciprocity are analyzed from symmetries satisfied by the superconductors. The proposed diode effect also applies to other chiral conductors as interlayer. Its formation has no restriction on the type of the current-phase difference relations. The generality and flexibility of our proposed diode effect provides more possibilities for the design of the dissipationless diode devices.

\end{abstract}
\maketitle

\section{\label{sec1}Introduction}

Semiconducting diodes are the essential units in the modern electronics.
As the typical nonreciprocal device, semiconducting diodes have
low resistances in one bias direction and high resistances in the other.
However, the energy consumption is unavoidable in the diodes
due to the finite resistance.
Therefore, realizing the superconducting diode effect becomes
a fundamental subject in condensed matter physics.
The superconducting diode possesses the different critical current
for the opposite current flow, which can achieve the
unidirectional nondissipative transport with zero resistance
in only one direction.

The nonreciprocal charge transport in superconductors can manifest itself
as the so-called magnetochiral anisotropy effect\cite{Rikken}
which requires the simultaneous breaking of the inversion and the time reversal symmetries\cite{Tokura}.
Experiments demonstrate the enhanced nonreciprocal coefficient characterizing the magnetochiral anisotropy in the noncentrosymmetric superconductors\cite{Wakatsuki1,Itahashi} and the topological insulator/superconductor interface\cite{Yasuda} for the temperature regime slightly above the critical value $T_{C}$. The theoretical researches suggest that the nonreciprocity of paraconductivity dominates in these systems for the temperature regime, which originates from the fluctuation of the superconducting order parameter\cite{Wakatsuki2,Hoshino}.

Recently, the nonreciprocity of supercurrent, i.e.,
the superconducting diode effect, has been observed in many experiments.
Ando et. al. demonstrate that the artificial superlattice without the inversion symmetry can host the magnetically controllable superconducting diode effect,
which exhibits zero resistance in only one direction\cite{Ando}.
The high rectification ratio comparable to the typical
semiconductor diodes can be obtained.
Lin et.al. find an intrinsic superconducting diode at zero external field
in mirror-symmetric twisted trilayer graphene,
which can be controlled by varying the carrier density or twist angle\cite{Lin}.
The physical origin is speculated as the formation of
the finite momentum Cooper pairs.
Lyu et.al. demonstrate a superconducting diode effect in the superconducting
film patterned with a conformal array of nanoscale holes,
in which the rectification DC voltage is measured
by applying the high frequency AC current\cite{Lyu}.
The rectification of the superconducting diode is also observed
in the NbSe$_{2}$ constrictions as a consequence of
the valley-Zeeman spin-orbit interaction\cite{Bauriedl}.
In addition, the nonreciprocity of supercurrent has been realized in some
Josephson junctions also.
For example, the Josephson diode, the superconducting diode
in the Josephson junctions, has been realized
in the inversion symmetry breaking van der Waals heterostructure
of NbSe$_2$/Nb$_3$Br$_8$/NbSe$_2$\cite{Wu}
and in highly transparent Josephson junctions fabricated on InAs quantum wells.\cite{Baumgartner}
As well, a giant Josephson diode effect is reported in Josephson junctions
formed on the type-\uppercase\expandafter{\romannumeral 2} Dirac semimetal\cite{Pal}.
The finite pairing momentum as its physical origin is established
from the evolution of the interference pattern.

Theoretical schemes for superconducting diode have also been proposed,
which involve both the bulk superconductors and the Josephson junctions\cite{Scammell,Hu,Zhang,Zinkl,Daido,Ilic,Yuan,Davydova,Zazunov,Reynoso,Chen,Dolcini,Kopasov,Alidoust,Minutillo,Halterman,Zhai,Legg,Yokoyama}. Very recent, Daido et.al. propose the intrinsic mechanism for
superconducting diode effect in noncentrosymmetric superconductors
by studying the nonreciprocity in the depairing current\cite{Daido}.
Ili$\acute{\text{c}}$ and Bergeret investigate the disorder effect
on superconducting diode caused by the helical superconducting state\cite{Ilic}.
Also, superconducting diodes are predicted in the polar film, few-layer MoTe$_2$,
the twisted bilayer graphene and the short Josephson junctions\cite{Yuan,Davydova}.
In these studies, the finite momentum pairing as the result of
the spin-orbit coupling\cite{Daido,Ilic,Yuan} or the Meissner
effect\cite{Davydova} is indispensable to realize superconducting diode.
In addition, the superconducting diode effect can also be achieved
based on the so-called $\phi_{0}$ Josephson junctions
which possess the anomalous current with the finite value at the zero phase difference\cite{Buzdin,Brunetti}.
For example, Josephson diode has been predicted
in the multilevel quantum dot\cite{Zazunov}, in the semiconductor nanowires\cite{Yokoyama}, in a spin polarizing quantum point contact in a two-dimensional electron gas\cite{Reynoso}, and in the topological insulator junctions\cite{Chen,Dolcini}.
In these studies based on the $\phi_{0}$ junctions, the spin-orbit coupling and an external field with fixed orientation are essential conditions to form
Josephson diodes.

In this work, we propose a concise Josephson diode composed of
two superconductors and a chiral quantum dot (QD) under an external magnetic field.
The distortion of the QD breaks the inversion symmetry
and the magnetic field is responsible for the time-reversal symmetry breaking.
Our proposed concise Josephson diode has the following three advantages:
The superconductors are of the plain $s$-wave paring and
do not need the formation of the finite momentum Cooper pairs.
The Josephson current hosted in our diode is not anomalous,
which has the zero current at the zero phase difference.
Furthermore, the spin-orbit coupling necessary for superconducting diode effects
in almost all existing researches is not required in our proposal.
The proposed Josephson diode also applies to other superconductors
such as the $d$-wave case and other chiral conductors\cite{Guo1,Guo2,Miyamoto}.
We use the Keldysh nonequilibrium Green's function formalism
with the self-consistent procedure to systemically study
the dependence of the supercurrent nonreciprocity in the diode
on various structure parameters such as the hopping amplitude,
the magnitude and orientation of the external field and the level of QD.
A strong nonreciprocity can be acquired in a large range of the parameter space.
The reversal and the switching of the on/off state of the diode can be easily regulated by inverting magnetic field and rotating it from the direction parallel to current to the perpendicular direction, respectively.
In addition, we also analyze the antisymmetry and the symmetry satisfied
by the nonreciprocity through the symmetric operations obeyed
by the conventional superconductors.

The organization of this paper is as follows.
In Sec.\uppercase\expandafter{\romannumeral 2}, we present the model of the proposed Josephson diode and the Keldysh nonequilibrium Green's function formalism for the calculations of the Josephson current.
In Sec.\uppercase\expandafter{\romannumeral 3}, We give the numerical results and discussions for the nonreciprocity of the Josephson current. The symmetric relations satisfied by the nonreciprocity are analyzed.
Sec. \uppercase\expandafter{\romannumeral 4} concludes this paper.

\section{\label{sec2}Model and formulation}

\begin{figure}[!htb]
\centerline{\includegraphics[width=0.8\columnwidth]{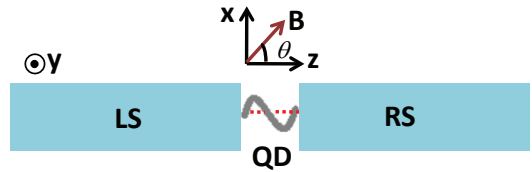}}
\caption{Schematic illustration of the superconducting diode proposed in this work. The left superconductor (LS) and right superconductor (RS) are separated by a chiral QD with distortion. The external field $\bf{B}$ is exerted on the QD with its polar angle $\theta$. The azimuthal angle $\varphi$ in the $xy$ plane is not shown here since the nonreciprocal effect in the diode is irrespective of it.}\label{fig1}
\end{figure}

Our proposed Josephson junctions are schematically shown in Fig.{\ref{fig1}},
which consist of the left superconductor, the right superconductor and a chiral QD under an external magnetic field.
The chiral QD can be experimentally fabricated and
can also be some natural chiral molecules,
such as the helical organic molecules\cite{Guo1,Guo2} or nanotubes\cite{Miyamoto}.
Notice that the junctions studied in this work based on
magnetochiral anisotropy possess different geometric configuration and operating principle from the theoretically proposed Josephson diodes based on the structural asymmetry\cite{Hu,Misaki}.
In Ref.[\onlinecite{Hu}], the different doping in two superconductors
facilitate the formation of a self-organized region,
which leads to the diode effect under the opposite bias voltages.
In Ref.[\onlinecite{Misaki}], the charging energy difference exists
in the two sides of the Josephson junctions,
which causes the nonreciprocity of the voltage drop.

The exerted external magnetic field in our junctions
is given by ${\bf{B}}=B(\sin{\theta}\cos{\varphi},\sin{\theta}\sin{\varphi},\cos{\theta})$ with its polar angle $\theta$ and the azimuthal angle $\varphi$ as shown in Fig.{\ref{fig1}}. The Josephson current $I$ can flow parallel to the $z$ axis, which will induce an extra magnetic field ${\bf{B}'}$ in QD due to its helicity. This extra field will be proportional to the current with the proportional coefficient $\alpha$. The Hamiltonian describing our model can be written as\cite{Sun1}
\begin{eqnarray}
H=H_L+H_{QD}+H_R+H_{T},
\end{eqnarray}
where $H_{L(R)}$, $H_{QD}$ and $H_T$ describe the left (right) superconductor, the central QD and the coupling between the superconductors and QD, respectively. They can be expressed as
\begin{eqnarray}
H_{L(R)}&=&\sum_{\bf{k(p)}\sigma}\epsilon_{\bf{k(p)}}c^{+}_{\bf{k(p)}\sigma}c_{\bf{k(p)}\sigma}
+\sum_{\bf{k(p)}}[\Delta_{L(R)}c_{\bf{k(p)}\downarrow}c_{-\bf{k(p)}\uparrow} \nonumber\\
&&+\Delta^{*}_{L(R)}c^{+}_{-\bf{k(p)}\uparrow}c^{+}_{\bf{k(p)}\downarrow}],
\label{LRSH}\\
H_{QD}&=&\sum_{\sigma}[(\epsilon_{d}+\sigma\alpha I+\sigma B\cos{\theta})d^{+}_{\sigma}d_{\sigma}\nonumber\\
&&+B\sin{\theta}e^{-i\sigma\varphi}d^{+}_{\sigma}d_{\bar{\sigma}}],
\label{QDH}\\
H_T&=&\sum_{\sigma}\left[\sum_{\bf{k}}v_{L}c^{+}_{\bf{k}\sigma}d_{\sigma}
 +\sum_{\bf{p}}v_{R}c^{+}_{\bf{p}\sigma}d_{\sigma}+\text{H.C.}\right].\label{HT}
\end{eqnarray}
Here, $\Delta_{L(R)}=\Delta_{L(R)0} e^{i\phi_{L(R)}}$ is the energy gap function
for the left (right) superconductor with the superconducting phase $\phi_{L(R)}$
and $v_{L(R)}$ is the hopping amplitude between the superconductors and QD.
$\epsilon_{d}$, $B$, and $I$ are the level of the QD, the magnitude of
the external magnetic field, and the current flowing through the junctions
which needs to be calculated self-consistently, respectively.

The Josephson current flowing in the junctions can be calculated
from the evolution of the number operator of the electrons in the left
superconductor $N_{L}=\sum_{\bf{k}\sigma}c^{+}_{\bf{k}\sigma}c_{\bf{k}\sigma}$(in units of $\hbar=1$),
\begin{eqnarray}
I=-e\left\langle  \frac{d}{dt}N_{L}(t)  \right\rangle.
\end{eqnarray}
After the Fourier transformation, it can be expressed as\cite{Sun2}
\begin{eqnarray}
I=2e\int\frac{dw}{2\pi}\sum_{\sigma}\text{Re}[v_{L}e^{i\phi_{L}}G^{<}_{dL\sigma\sigma,11}(w)],\label{current}
\end{eqnarray}
where $G^{<}_{dL\sigma\sigma,11}(w)$ is the $(1,1)$ element of
the Fourier transformation of the Keldysh Green's function $G^{<}_{dL\sigma\sigma}(t)$. We have defined
\begin{eqnarray}
G^{<}_{\beta\beta'\sigma\sigma'}(t)&=&
i\sum_{\bf{q}\bf{q'}}\left\langle
\begin{pmatrix}
c^{+}_{\bf{q}'\sigma'}(0),c_{-\bf{q}'\bar{\sigma'}}(0)
\end{pmatrix}
\otimes
\begin{pmatrix}
c_{\bf{q}\sigma}(t)\\
c^{+}_{-\bf{q}\bar{\sigma}}(t)
\end{pmatrix}
\right\rangle,\label{Gbb}\\
G^{<}_{d\beta\sigma\sigma'}(t)&=&i\sum_{\bf{q}}\left\langle
\begin{pmatrix}
c^{+}_{\bf{q}\sigma'}(0),c_{-\bf{q}\bar{\sigma'}}(0)
\end{pmatrix}
\otimes
\begin{pmatrix}
d_{\sigma}(t)\\
d^{+}_{\bar{\sigma}}(t)
\end{pmatrix}
\right\rangle,\\
G^{<}_{dd\sigma\sigma'}(t)&=&i\left\langle
\begin{pmatrix}
d^{+}_{\sigma'}(0),d_{\bar{\sigma'}}(0)
\end{pmatrix}
\otimes
\begin{pmatrix}
d_{\sigma}(t)\\
d^{+}_{\bar{\sigma}}(t)
\end{pmatrix}
\right\rangle,
\end{eqnarray}
with $(\beta,{\bf{q}})$ or $(\beta',\bf{q}')$ being $(L,{\bf{k}})$ for
the left superconductor and being $(R,{\bf{p}})$ for right superconductor,
which are $2\times2$ matrices in the particle-hole space.
They form the Keldysh Green's function $G^{<}$ defined in the space spanned
by the three regions: the left and right superconductors and QD,
\begin{eqnarray}
G^{<}=\left(\begin{array}{ccc}
G_{LL}^{<}&G_{LR}^{<}&G_{Ld}^{<}\\
G_{RL}^{<}&G_{RR}^{<}&G_{Rd}^{<}\\
G_{dL}^{<}&G_{dR}^{<}&G_{dd}^{<}
\end{array}\right),
\end{eqnarray}
with its elements are $4\times4$ matrices in the spin$\otimes$particle-hole space. For example, $G_{LL}^{<}$ can be written as
\begin{eqnarray}
G_{LL}^{<}=\left(\begin{array}{cc}
G_{LL\uparrow\uparrow}^{<}&G_{LL\uparrow\downarrow}^{<}\\
G_{LL\downarrow\uparrow}^{<}&G_{LL\downarrow\downarrow}^{<}
\end{array}\right),
\end{eqnarray}
with $G_{LL\sigma\sigma'}^{<}$ defined in Eq.(\ref{Gbb}).

The Green's function $G^{<}$ can be solved from the fluctuation-dissipation theorem
\begin{eqnarray}
G^{<}(w)=-f(w)(G^{r}-G^{a}),\label{Keldysh}
\end{eqnarray}
with the Fermi distribution function $f(w)$.
Here $G^{r}$ and $G^{a}$ are the corresponding retarded Green's function and the advanced Green's function. The former satisfies the Dyson equation
\begin{eqnarray}
G^{r}=g^r+g^{r}\Sigma^{r}G^{r},
\end{eqnarray}
and $G^a$ is its Hermitian conjugation. The self energy $\Sigma^{r}$ is given by
\begin{eqnarray}
\Sigma^{r}=\left(\begin{array}{ccc}
0&0&V_{L}\\
0&0&V_{R}\\
V^{*}_{L}&V_{R}^{*}&0
\end{array}\right),
\end{eqnarray}
with $V_{L(R)}=1_{2\times2}\otimes\text{diag}(v_{L(R)}e^{i\phi_{L(R)}/2},-v^{*}_{L(R)}e^{-i\phi_{L(R)}/2})$.

The retarded Green's function $g^{r}$ of the system without coupling between the superconductors and QD can be obtained exactly as\cite{Sun3}
\begin{eqnarray}
g^{r}=\text{diag}(g^{r}_{LL},g^{r}_{RR},g^{r}_{dd}),
\end{eqnarray}
in which $g^{r}_{LL(RR)}=\text{diag}(g^{r}_{LL(RR)\uparrow\uparrow},g^{r}_{LL(RR)\downarrow\downarrow})$ with
\begin{eqnarray}
g^{r}_{LL(RR)\sigma\sigma}=-i\pi\rho_{N}\rho_{L(R)}\left(\begin{array}{cc}
1&\frac{\sigma\Delta_{L(R)0}}{w+i0^{+}}\\
\frac{\sigma\Delta_{L(R)0}}{w+i0^{+}}&1
\end{array}\right),
\end{eqnarray}
for the isolated left (right) superconductor.
Here, the dimensionless BCS density of states normalized by the normal density of states $\rho_{N}$ is given by $\rho_{L(R)}=\vert w\vert/\sqrt{w^2-\Delta_{L(R)0}^2}$ for $\vert w\vert>\Delta_{L(R)0}$ and $-iw/\sqrt{\Delta_{L(R)0}^2-w^2}$ for $\vert w\vert<\Delta_{L(R)0}$. The retarded Green's function for the isolated QD is given by
\begin{eqnarray}
g^{r}_{dd}=\left(\begin{array}{cc}
g^{r}_{dd\uparrow\uparrow}&g^{r}_{dd\uparrow\downarrow}\\
g^{r}_{dd\downarrow\uparrow}&g^{r}_{dd\downarrow\downarrow}
\end{array}\right),\label{grdd}
\end{eqnarray}
with $g^{r}_{dd\sigma\sigma}=\text{diag}(L_{\bar{\sigma}-}/(L_{\sigma-}L_{\bar{\sigma}-}-\vert B_{\sigma}\vert^2),L_{\bar{\sigma}+}/(L_{\sigma+}L_{\bar{\sigma}+}-\vert B_{\sigma}\vert^2))$, $g^{r}_{dd\sigma\bar{\sigma}}=\text{diag}(B_{\bar{\sigma}}/(L_{\sigma-}L_{\bar{\sigma}-}-\vert B_{\sigma}\vert^2),-B_{\bar{\sigma}}/(L_{\sigma+}L_{\bar{\sigma}+}-\vert B_{\sigma}\vert^2))$. Here, $L_{\sigma\pm}=(w+i0^{+}\pm\epsilon_{d})-\sigma\alpha I-\sigma B\cos{\theta}$ and $B_{\sigma}=B\sin{\theta}e^{i\sigma\varphi}$.

After solving the solution of $G^{<}$ from Eq.(\ref{Keldysh}), the Josephson current in Eq.(\ref{current}) can be calculated self-consistently\cite{Sun4}. Note that $G^{<}$ is also a function of the current $I$. In the following calculations, we introduce $\phi=\phi_{L}-\phi_{R}$ to denote the superconducting phase difference. We assume that the two superconductors are of the same material, which may be easier to implement experimentally. Therefore, the left and right superconductors
will possess the same energy gap, i.e., $\Delta_{L0}=\Delta_{R0}$. In this situation, it is reasonable to assume the symmetric hopping amplitude, i.e., $v_{L}=v_{R}$. Next, we will use $\Delta_{0}$ and $v$ to denote them in our calculations. The units of $v$, $B$ and $\epsilon_{d}$ will be chosen as $\Delta_{0}$. The normal density of states $\rho_{N}=1$ is adopted. For definiteness, the proportional coefficient $\alpha$ in front $I$ in Eq.({\ref{QDH}}) is taken as $0.05$.

\section{\label{sec3}Numerical results}
\begin{figure}[!htb]
\centerline{\includegraphics[width=0.8\columnwidth]{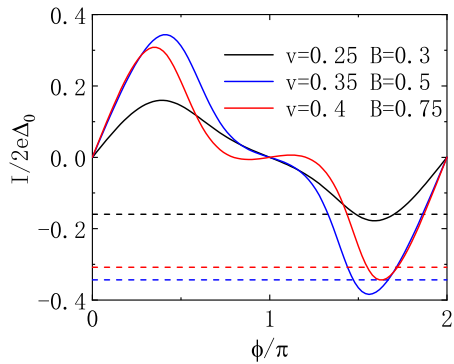}}
\caption{The current-phase difference relations for the given hopping amplitude $v$ and the external magnetic field $B$. The black, blue and red dashed lines are the critical current $-I_{c+}$ for $(v,B)=(0.25,0.3),(0.35,0.5)$ and $(0.4,0.75)$, respectively. They intersect rather than being tangent with the current in $\pi<\phi<2\pi$ due to the nonreciprocal effect.
The other parameters are taken as $(\epsilon_{d},\theta)=(0,0)$.}
\label{fig2}
\end{figure}

We plot the current-phase difference relations with
the QD's level $\epsilon_{d}=0$ and the magnetic field's polar angle $\theta=0$ in Fig.{\ref{fig2}}.
In this situation, the direction of the external magnetic field is along the $+z$ axis. It is found that the critical value for the positive current in $0<\phi<\pi$ is different from that for the negative current in $\pi<\phi<2\pi$. We define the former critical value as $I_{c+}$ ($I_{c+}=\text{max}[I(0<\phi<2\pi)]$) and the latter as $I_{c-}$ ($I_{c-}=\text{max} [-I(0<\phi<2\pi)]$) and one has $I_{c-}>I_{c+}$.
This is just the nonreciprocal effect in our junctions, which can be seen obviously in the figure. If we plot $-I_{c+}$ with given parameters $v$ and $B$ using the dashed lines, they will intersect the negative current with the same parameters. This is different from the reciprocal current-phase difference relations with $I_{c+}=I_{c-}$ where the dashed lines will be tangent with the negative current. Due to the difference between $I_{c+}$ and $I_{c-}$, the junctions can be used as a highly effective rectifier by changing the direction of the applied current $I$ with $I_{c+}<I<I_{c-}$. For the positive current, the junctions become the normal state with finite resistance, while for the negative current, the junctions turn into the superconducting state with zero resistance. This type of rectification has been recently observed experimentally in the superconducting diode based on the artificial superlattice\cite{Ando} or the small-twist-angle trilayer graphene\cite{Lin} and the Josephson diode based on the van der Waals heterostructure\cite{Wu}.

There are two other remarkable characters in the current-phase difference relations in Fig.{\ref{fig2}}.
The first is that the supercurrent is equal to zero at $\phi=n\pi$ with $n$ an integer number.
That is, there is no anomalous supercurrent in our junctions.
The second is that the critical values of the current are obtained at the symmetric positions with respect to $\phi=\pi$ in the range $0<\phi<2\pi$ or with respect to $\phi=0$ in the range $-\pi<\phi<\pi$. Actually, the former character can survive in the whole space of parameters while the latter is almost exactly true for small $\alpha$ and holds approximately for larger $\alpha$. In a word, our junctions basically keep the characters of $\sin{\phi}$ which is just the initial form of current in the junctions without the chirality of QD.
The diode effect is caused simply by the enhancement of the initial critical current for one direction and the reduction for the other.

Therefore, we can conclude that the occurrence of the nonreciprocity
in our schemes will not impose restrictions on the type of the initial current.
One can obtain various diode effects with the basic characters of a single harmonic. If we substitute one of the superconductors for a spin-triplet one, the initial current of the $\cos{\phi}$-type or the $\sin{2\phi}$-type can be realized due to the presence of the external magnetic field before introducing the chirality\cite{Brydon,Cheng1}.
It can be expected that the diode effect with the basic characters of $\cos{\phi}$ or $\sin{2\phi}$ will be obtained after the chirality is introduced in QD.
This is distinct from the nonreciprocity based on the $\phi_{0}$ junctions with the finite current at $\phi=0$.\cite{Reynoso,Zazunov}
There, the form of the initial current is changed fundamentally by the insertion of the intermediate layer which is used to introduce new harmonics.
Usually, the formation of the nonreciprocity in this situation need the combination of $\sin{\phi}$, $\cos{\phi}$ and higher order sinusoidal terms such as $\sin{2\phi}.$\cite{Baumgartner}
Of cause, the diode effect based on the $\phi_{0}$ junctions can also be realized in our junctions if one replaces the both superconductors by the helical $p$-wave ones\cite{Cheng2}.

In order to quantitatively characterize the nonreciprocity in our junctions,
we introduce the Josephson diode efficiency $\eta$ defined as
\begin{eqnarray}
\eta=\frac{I_{c-}-I_{c+}}{I_{0}},\label{eta}
\end{eqnarray}
with $I_{0}=(I_{c+}+I_{c-})/2$. For the current-phase difference relations in Fig.{\ref{fig2}},
the diode efficiency $\eta$ can exceed $10\%$ which indicate that our proposed superconducting diode can provide high nonreciprocity.

\begin{figure}[!htb]
\centerline{\includegraphics[width=\columnwidth]{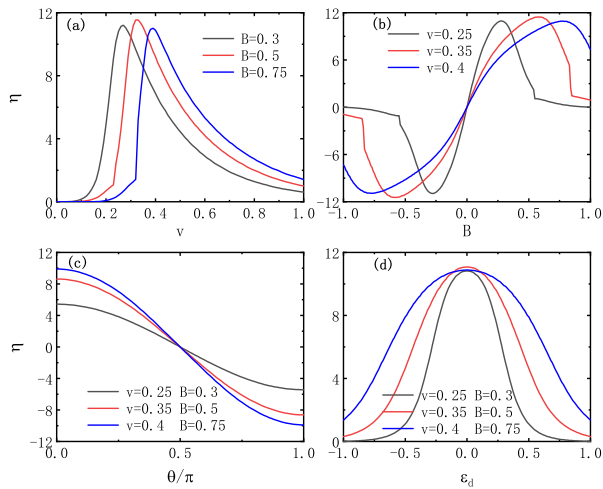}}
\caption{The evolutions of the nonreciprocity with the changes of (a) the hopping amplitude, (b) the magnitude of the external magnetic field, (c) the polar angle of the magnetic field and (d) the level of QD. The other parameters are taken as $(\epsilon_{d},\theta)=(0,0)$ in (a) and (b), $\epsilon_{d}=0.3$ in (c) and $\theta=0$ in (d).}
\label{fig3}
\end{figure}

Fig.{\ref{fig3}} shows the evolutions of the nonreciprocity when the parameters
are changed. The main features of the diode efficiency $\eta$
can be summarized as follows.
(1) There is a peak in the $\eta$ curve with the variation of
the hopping amplitude $v$ for a given magnetic field $B$ as shown in Fig.{\ref{fig3}}(a). The value of $v$ for the peak becomes larger when $B$ is increased. As $v$ tends to $0$, $\eta$ quickly drops to zero.
(2) The diode efficiency $\eta$ curve is antisymmetric
about the magnetic field $B=0$ as shown in Fig. {\ref{fig3}}(b) for a given $v$.
The value of $B$ for the peak of $\eta$ also becomes larger when $v$ is increased.
(3) The $\eta$ curve is antisymmetric about the polar angle $\theta=0.5\pi$
of the magnetic field for given $v$ and $B$ as shown in Fig.{\ref{fig3}}(c).
As the polar angle $\theta$ tends to $0$ or $\pi$, the nonreciprocity will be enhanced.
(4) The $\eta$ curve is symmetric about the QD's level $\epsilon_{d}=0$ as shown in Fig.{\ref{fig3}}(d). The peak value of $\eta$ for given $v$ and $B$ is obtained at $\epsilon_{d}=0$, i.e., the level $\epsilon_d$ lying in the middle of the energy gaps of the left and right superconductors.
(5) The nonreciprocity characterized by $\eta$ is irrespective of the azimuthal angle $\varphi$ of the external magnetic field. Therefore, the parameter $\varphi$ does not appear in our numerical results.

We now discuss the physical origin of the nonreciprocity in our setup,
which is responsible for the main features.
The nonreciprocity in the junctions originates from the different total field felt
by electrons in QD when the direction of the current is reversed.
Now, we assume that the external magnetic field is fixed along the $+z$ axis. Following the Hamiltonian $H_{QD}$ in Eq.(\ref{QDH}), the total field felt by the spin-up/down electrons is
\begin{eqnarray}
\pm(B+\alpha\vert I\vert),
\label{pf}
\end{eqnarray}
for the positive current while the total field felt by the spin-up/down electrons becomes
\begin{eqnarray}
\pm(B-\alpha\vert I\vert),
\label{nf}
\end{eqnarray}
for the negative current.
Here, $\alpha\vert I\vert$ is the magnitude of the extra field ${\bf{B}'}$
induced by the current $\vert I\vert$ in the chiral QD.
The opposite sign in front of $\alpha\vert I\vert$ for the positive
and the negative current comes from the chirality of QD.
In other words, the positive current will experience the geometry
with the opposite helicity compared with the negative current.
The opposite chiral current in QD will induce the extra field ${\bf{B}'}$
with the opposite direction. The superposition of ${\bf{B}}$ and ${\bf{B}'}$
will lead to the total field in Eqs.({\ref{pf}}) and ({\ref{nf}}).

For the phase difference $\phi=n\pi$, the initial current of
the $\sin{\phi}$-type before introducing the chirality in QD is zero.
Then the induced field ${\bf{B}'}$ are vanished.
So the current is still zero after the chirality is introduced in QD.
On the other hand, while $\phi \not= n\pi$,
the critical value of the initial current can induce
the largest magnitude of $\bf{B}'$ and hence the nonreciprocity.
The maximum and the minimum of the current for the presence of chirality in QD
will be located around their initial positions under the self-consistent calculation. That is why our junctions can keep the basic characters of
the initial current as shown in Fig.{\ref{fig1}}.

From Eqs.({\ref{pf}}) and (\ref{nf}), one can find that the external magnetic field and the chirality of QD are necessary conditions for the nonreciprocity. If ${\bf{B}}=0$, Eqs.({\ref{pf}}) and $(\ref{nf})$ turn into $\pm\alpha\vert I\vert$ and $\mp\alpha\vert I\vert$, respectively. Taking the contributions from the two spins into account, the critical values for the positive and the negative currents will not be different. This is the reason for the diode efficiency $\eta=0$ at $B=0$ in Fig.\ref{fig3}(b).
If the chirality is absent in QD, $\alpha\vert I\vert$ in Eqs.({\ref{pf}}) and (\ref{nf}) will disappear. The total magnetic field felt by carriers
for the positive current and the negative current also keep the same form,
which will not bring the nonreciprocal effect.
When both the external field $B$ and the chirality are present in QD and
the external field is increased, larger value of $\alpha\vert I\vert$ is needed to observe the obvious nonreciprocal effect.
For a given $\alpha$, this can be achieved by raising the current $I$,
which is equivalent to raising the hopping amplitude $v$ or keeping the level of QD $\epsilon_{d}=0$. That is why the peak of $\eta$ moves right with the increasing $B$ from $0.25$ to $0.4$ in Fig.{\ref{fig3}}(a) and the peaks will be always located at $\epsilon_{d}=0$ in Fig.{\ref{fig3}}(d). In contrast, decreasing $v$ or deviating the level of QD from $\epsilon_{d}=0$ will weaken $\eta$ as shown in Figs.{\ref{fig3}}(a) and (d). But the excessive hopping amplitude will reduce the value of $\eta$ as shown in Fig.{\ref{fig3}}(a) since it causes a rapid increase of current, including its critical value $I_{0}$ as the denominator of $\eta$ in Eq.({\ref{eta}}).

From Eqs.({\ref{pf}}) and (\ref{nf}), we can also find that the inversion of the external field ${\bf{B}}$ will invert the nonreciprocity.
Now, we fix the external field to along the $-z$ axis, i.e., $B\rightarrow-B$. Eqs.({\ref{pf}}) and (\ref{nf}) will become $\mp(B-\alpha\vert I\vert)$ and $\mp(B+\alpha\vert I\vert)$, respectively. The total field felt by electrons for the positive current and the negative current is exchanged. The critical current $I_{c+}(I_{c-})$ becomes $I_{c-}(I_{c+})$ and $\eta$ will change its sign according to its definition in Eq.(\ref{eta}), which will give rise to the antisymmetry of $\eta$ about $B=0$ as shown in Fig.{\ref{fig3}}(b).
For the external field not parallel to the $z$ axis, Eqs. ({\ref{pf}}) and ({\ref{nf}}) will be replaced by $\pm(B\cos{\theta}+\alpha\vert I\vert)$ and $\pm(B\cos{\theta}-\alpha\vert I\vert)$ (see Eq.({\ref{QDH}})), respectively. According to the above discussions, if the external field is perpendicular to the current with $\theta=0.5\pi$, $B\cos{\theta}=0$ and the nonreciprocity will be vanished. The polar angles $\theta$ and $\pi-\theta$ will bring about the inverse sign of $B\cos{\theta}$ and hence the inverse nonreciprocity, i.e., $\eta(\theta)=-\eta(\pi-\theta)$, which is explicitly shown in Fig.{\ref{fig3}}(c). Consequently, one can conveniently control the inversion and the on/off state of the nonreciprocity of the junctions by tuning the direction of the external field.

It is worth to note that the antisymmetric property of the nonreciprocity
about the external magnetic field in our Josephson junctions based on the conventional superconductors is consistent with the nonreciprocal effect
observed in the bulk Rashba superconductor\cite{Ando}.
But the nonreciprocity observed in the van der Waals heterostructure possesses
the symmetric property about the external field because its mechanism may be the asymmetric tunneling induced by polarization rather than the magnetochiral anisotropy\cite{Wu}. Moreover, the dependence of the nonreciprocity in our junctions on the orientation of the external field is distinct from the Josephson diode effect based on the type-$\text{\uppercase\expandafter{\romannumeral2}}$ Dirac semimetal with the mechanism of the finite Cooper pair momentum\cite{Pal}.
There, the nonreciprocity obtains its largest value when the in-plane field is perpendicular to the current and vanishes when the field is parallel to the current. This property is contrary to the nonreciprocal effect here in our Josephson junctions.

The properties of the diode efficiency $\eta$ can also be analyzed from the perspective of symmetries obeyed by superconductors\cite{Cheng3,Cheng4,Cheng5}. First, the superconductor Hamiltonians $H_{L}$ and $H_{R}$
satisfy the symmetry of the spin-rotation about the $z$ axis.
If we use $\mathcal{R}_{z}(\gamma)$ to denote the $\gamma$-angle rotation,
$H_{QD}$ in Eq.(\ref{QDH}) under this unitary transformation will become $\mathcal{R}_{z}(\gamma)H_{QD}(\theta,\varphi)\mathcal{R}_{z}(\gamma)^{-1}=H_{QD}(\theta,\varphi+\gamma)$. Since the Josephson current is invariant under the unitary transformation, $I(\theta,\varphi)=I(\theta,\varphi+\gamma)$ must be satisfied. This equality indicates that the Josephson current and hence the diode efficiency $\eta$ are irrespective of the azimuthal angle $\varphi$. Second, the time-reversal operation $\mathcal{T}$ can only change the superconducting phase difference from $\phi$ to $-\phi$. Simultaneously, we have $\mathcal{T}H_{QD}(I,\theta,\varphi)\mathcal{T}^{-1}=H_{QD}(-I,\pi-\theta,\varphi+\pi)$. Since $\mathcal{T}$ will reverse the Josephson current, one obtains $I(\theta,\varphi)=-I(\pi-\theta,\pi+\varphi)$ and hence $\eta(\theta,\varphi)=-\eta(\pi-\theta,\varphi+\pi)$, i.e., $\eta({\bf{B}})=-\eta(-{\bf{B}})$. Taking the independence of $\eta$ on $\varphi$ into account, the antisymmetric relation $\eta(\theta)=-\eta(\pi-\theta)$ will be derived. Third, the left and right superconductors also keep invariant under the unitary joint transformation $\mathcal{S}\mathcal{U}(\pi/2)$. Here, $\mathcal{S}c_{\bf{k}(\bf{p})\uparrow}\mathcal{S}^{-1}=c^{+}_{-\bf{k}(\bf{p})\downarrow}$, $\mathcal{S}c_{\bf{k}(\bf{p})\downarrow}\mathcal{S}^{-1}=-c^{+}_{-\bf{k}(\bf{p})\uparrow}$ and $\mathcal{U}(\pi/2)$ is the $U_{1}$ gauge transformation with the phase $\pi/2$. $H_{QD}(\epsilon_{d})$ under the same transformation turns into $H_{QD}(-\epsilon_{d})$, which means the current satisfies $I({\epsilon_d})=I(-\epsilon_{d})$. Therefore, the diode efficiency $\eta$ is symmetric about $\epsilon_d=0$ as shown in Fig.{\ref{fig3}}(d).

\begin{figure}[!htb]
\centerline{\includegraphics[width=\columnwidth]{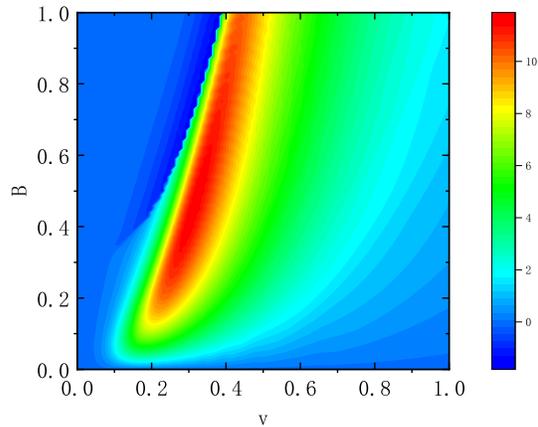}}
\caption{The Josephson diode efficiency $\eta$ versus the hopping amplitude $v$
and the external magnetic field $B$.
The other parameters are take as $(\epsilon_{d},\theta)=(0,0)$.
The colour bar denotes its percentage.}
\label{fig4}
\end{figure}

In order to show a complete picture of the nonreciprocity, we present the contour plot of the diode efficiency $\eta$ with $(\epsilon_{d},\theta)=(0,0)$ in the $(v,B)$ space in Fig.{\ref{fig4}}.
The dependences of $\eta$ on the hopping amplitude $v$ and the field $B$
can be seen more directly. In a large area of the space, $\eta$ can achieve its peak value exceeding $10\%$. The optimal hopping amplitude roughly distributes in the interval of $0.2<v<0.5$. In this interval, the nonreciprocity for the small field with $B=0.2$ can also reach about $10\%$. Actually, even for $B=0.05$, the nonreciprocity of $5\%$ can still be obtained. In the calculations, we have assumed $v_{L}=v_{R}$ and $\Delta_{L}=\Delta_{R}$. However, it should be emphasized that the mismatch of the energy gaps or the hopping amplitudes will not fundamentally change the superconducting diode effect in the proposed junctions.
The presence of the intradot $e$-$e$ Coulomb interaction will also not ruin the nonreciprocal effect. These properties prove the robustness of the nonreciprocity in our junctions and its realizability in experiment.

Finally, we give an estimation of the value of $\alpha$. We model the QD as a helix with a single pitch of $2nm$ and the radius $R=1nm$. In order to simply calculate the induced field $B'$, the helix is approximately taken as a coil carrying current $I$ with the same radius $R$. Using the Biot-Savart law, we can solve $\alpha=g\mu_{B}\mu_{0}/4R$ with the Bohr magneton $\mu_{B}$ and the vacuum permeability $\mu_{0}$. If we take the Lande factor $g=2$ and a single level in QD into account, the magnitude of $\alpha$ is about $1.8\times 10^{-5}$. Here, the unit of $\alpha$ is $\hbar/2e$ since the unit of current has been chosen as $2e\Delta_{0}/\hbar$. For larger Lande factor and more modes of QD in the $x$ or $y$ direction, the magnitude about $10^{-2}$ of $\alpha$ can be obtained. So it is possible to take $\alpha=0.05$ in our numerical calculations. For smaller value $\alpha=0.01$, the nonreciprocity exceeding $5\%$ is still can be obtained for optimal parameters.

\section{\label{sec5}Conclusions}
In summary, the Josephson junctions consisting of the conventional superconductors and a chiral quantum dot (QD) can host the nonreciprocity of supercurrent,
which show outstanding universality and flexibility.
The formation of the diode effect does not require spin-orbit coupling
to cause the finite momentum of Cooper pairs or change the initial current-phase difference relations.
It applies to superconductors with different pairings,
various of chiral conductors and arbitrary current-phase difference relations.
The inversion and the on/off state switching of the nonreciprocity in the diode
can be easily controlled by tuning the direction of the external magnetic field.
The superconducting diode effect strongly depends on the junction parameters.
The strong nonreciprocity can be obtained in a large area of the parameter space composed of the hopping amplitude and the external field.
The symmetry of the nonreciprocity about the level of QD and its antisymmetry about the reversal of the external field are derived, which are analyzed from the symmetric operations obeyed by the superconductors. Our researches possess important theoretical and practical values for the exploration of the nonreciprocal devices with no dissipation.

\section*{\label{sec5}ACKNOWLEDGMENTS}
This work was financially supported by NSF-China under Grants Nos. 11921005 and 11447175, the Strategic Priority Research Program of Chinese Academy of Sciences (XDB28000000), and the Natural Science Foundation of Shandong Province under Grants No. ZR2017QA009.

\end{document}